# Contention Based Routing in Mobile Ad Hoc Networks with Multiple Copies

Ms. E. Jenefa JebaJothi , Dr. V. Kavitha , Ms. T. Kavitha

**Abstract —** Routing the packets efficiently in mobile ad hoc network does not have end to end paths. Multiple copies are forwarded from the source to the destination. To deal with such networks, researches introduced flooding based routing schemes which leads to high probability of delivery. But the flooding based routing schemes suffered with contention and large delays. Here the proposed protocol "Spray Select Focus", sprays a few message copies into the network, neighbors receives a copy and by that relay nodes we are choosing the shortest route and then route that copy towards the destination. Previous works assumption is that there is no contention and dead ends. But we argue that contention and dead ends must be considered for finding efficiency in routing. So we are including a network which has contention and dead ends and we applied the proposed protocol. We can say that this protocol works well for the contention based network.

**Index Terms**—Ad hoc, contention, deadend and Routing.

——————————— ◆ ———————————

## 1 INTRODUCTION

Routing efficiently in a mobile ad hoc network does not have end to end path from the source to the destination. The concept of connected, stable network over which data can be routed reliably rarely holds there. In case of wireless signals are subject to multi-path propagation, fading and interference making wireless links unstable and lossy [1]. Additionally, [6] frequent node mobility significantly reduces the time a good link exists and constantly changes the network connectivity graph. As a result wireless connectivity is volatile and usually intermittent and complete end-to-end paths will not exist [1]. Tactical networks may also choose to operate in an intermittent fashion for Low probability of interception and low probability of detection. Deep space networks and underwater networks often have to deal with parameters such as long propagation delays and or intermittent connectivity, as well [8] [9]. These networks are referred to as Delay tolerant networks [10].

These networks can neither make any assumptions about the existence of a contemporaneous path to the destination nor assume accurate knowledge of the destination's location or even addresses.

Under such intermittent connectivity or networks conditions many traditional protocols fail. The biggest challenge is that to enable networking in intermittently connected or mobile network environment is routing.

————————————————

- E.Jenefa JebaJothi is with the Department of Computer Science and Engineering, Anna University Tirunelveli, Tirunelveli, Tamilnadu.
- Dr. V. Kavitha is with the Department of Computer Science and Engineering, Anna University Tirunelveli, Tirunelveli, Tamilnadu.
- T.Kavitha is with the Department of Computer Science and Engineering, Anna University Tirunelveli, Tirunelveli, Tamilnadu.

Conventional internet routing protocols as well as routing schemes for mobile ad hoc networks assume that a complete path exists between a source and a destination and try to discover these paths before any useful data is sent. Thus if no end-to-end paths exist most of the time, these protocols fail to deliver any data to all but the few connected nodes.

However this does not mean that packets can never be delivered in these networks. In mobility assisted routing [6] a message could be sent over an existing link, get buffered at the next hop until the next link in the path comes up and so on and so forth, until it reaches its destination. The utility-based flooding scheme is quite fast in some scenarios, the overhead involved in terms of bandwidth, buffer space, and energy dissipation is often prohibitive for small wireless devices. In multi-copy scheme, more than one copy per message was used and in single-copy scheme only route one copy per message can considerably reduce resource waste. So no routing scheme for intermittently connected environments currently exists that can achieve both small delays and prudent usage of the network and node resources.

The problem of contention in the network and dead ends are not concentrated in the previous works. But we say that contention and dead ends are important factors to be considered. Ignoring contention and dead ends will give inaccurate results [9].

For this reason the implementation of multi-copy protocols called Spray routing is introduced [2] by considering contention and dead ends in the network.

## 2 RELATED WORK

Mobile ad hoc's do not have complete end-to-end path. So an opportunistic hop-by-hop routing is used. The first work done with a single copy of message sprayed in the network [11]. It uses the Single copy routing algorithm which says, to reach the destination node, the current node holding the single message copy will handover the message to another node it encounters. If it does not have a path or relay it performs direct transmission in which a node a forwards a message to another node B it encounters, only if b is the messages' destination. The above algorithm has bad transmission rate when the single copy get lost and has good delivery delay. Multiple copies of messages for transmission lead to flooding. Spray and Wait routing algorithm is used here.

Second work Sprays a multiple number of copies [1] into the network and then waits till one of these nodes meets the destination. It contains an algorithm called spray and wait with which we have taken for analysis. There are two phases Spray phase which spreads the copies and Wait phase which performs direct transmissions. This routing scheme is highly scalable and has reasonable delays.

Third related work with a routing scheme called Binary Spray and Wait routing algorithm [10] works as, the source of a message initially starts with L copies; any node A that has n>1 message copies, encounters another node B with no copies, hands over to B, n/2 and keeps n/2 for itself; when it is left with only one copy, it switches to direct transmission. This algorithm performs well in both **message** delivery and transmissions rate.

The fourth scheme is similar to the single copy routing scheme which scheme uses only one copy per message. Seek and Focus (hybrid) routing algorithm [2] is used. Here each node maintains a timer for every other node. Nodes emit beacon signal, which advertise their presence. Other nodes which sense this beacon signal and establish a relationship by exchange id, called encounter. A node holding the single message copy, will handover to another node it encounters. The above algorithm has bad transmission rate when the single copy get lost.

The fifth scheme uses multiple copies and it uses Spray and Focus scheme, here multiple copies are sprayed into the network and focused to the destination by utility values [1] [6]. There is inaccuracy, because this paper assumed that there is no contention and dead ends in network. So contention and dead ends are important issues for accuracy. The following papers justifies that we must consider contention and dead ends.

Epidemic routing is a robust transmission scheme for ad hoc networks. Under the assumption of no contention, it has the minimum end-to-end delay amongst all the routing schemes. The assumption of no contention was justified by arguing that since the network is sparse, there will be very few simultaneous transmissions [1] [10]. Through simulations authors [9] proved that this argument is not correct and that contention cannot be ignored while analyzing the performance of routing schemes, even in sparse networks.

A large body of work has theoretically analyzed the performance of mobility-assisted routing schemes for intermittently connected mobile networks. However majority of these studies have ignored wireless contention. Previous works shown through simulations that ignoring contention leads to inaccurate and misleading results. Here the authors optimized the routing schemes using analytical expressions and computed expected delays which ignore contention and lead to suboptimal or even erroneous behavior.

## 3 ROUTING

In this section, we explore the problem of efficient routing in mobile ad hoc networks, and describe our proposed solution, Spray Select Focus Routing. Our problem setup consists of a number of nodes moving inside a bounded area according to a stochastic mobility model. Additionally, we assume that the network is disconnected at most times, and that transmissions are faster than node movement.

Our study of single-copy routing algorithms [11] showed that using only one copy per message is often not enough to deliver a message with high reliability and relatively small delay. At the same time, routing too many copies in parallel, as in the case of epidemic routing, can often have disastrous effects on performance. Flooding based schemes begin to suffer severely from contention as traffic increases, and their delay increases rapidly. Based on the above observations, we have identified the following goals for a routing protocol in mobile ad hoc networks:

- *Perform significantly fewer transmissions than flooding-based schemes.*

- *Deliver a message faster than existing schemes with optimal delays.*

- *Highly scalable*

- *Simple*

### 2.1 Spray Select Focus Routing

Although Spray and Focus [6] routing performs well in some scenarios, we say it is inaccurate because of not considering the contention and dead ends in the network. In previous works [8] [9], the authors argue that the performance of a routing scheme is accurate when it is subjected to contention and dead ends in the network.

### 2.2 Contention

Contention means competition for resources. Contention is defined as that two or more nodes may try to send messages across the network simultaneously. In Spray and Focus algorithm the contention is not considered and they optimized the copies. As per [7] [8], for congestion adaptive routing the path is minimized for routing. In our algorithm we spray multiple copies to the neighbors from that neighbors we are finding a route; if the copy is reached we will discard the other copies. Other wise we see the other copies for transmission. So we are minimizing the route to avoid contention in the network. Our routing algorithm has three phases:

*Spray*

For every message originating at a source node, L message copies are initially spread-forwarded by the source and possibly other nodes receiving a copy-to L distinct relays.

*Select*

Selects a node; from that node find the shortest route by hop distances to the destination.

*Focus*

Let $U_x(Y)$ denote the utility of node X for destination Y; a node A carrying a copy for destination D, forwards its copy to a new node B it encounters, if and only if $U_B(D) > U_A(D)$.

By using the Spray Select Focus Algorithm we are reducing the path and not the copies.

ALGORITHM FOR SPRAY SELECT FOCUS

- **Spray**
  1. Spray the message copies from the source
  2. Check for coverage
  3. If there is coverage
  4. Nodes which are in neighborhood receives a copy
- **Select**
  5. Nodes visited must not be visited again
  6. Minimization of route is done
  7. Copy is forwarded to the destination.
- **Focus**
  8. If the destination is not found
  9. Let A be a node having copy for destination D
  10. A forwards the copy to a new node B
      If $U_B(D) > U_A(D)$

## 2.3 DeadEnd

Occur when the node gets struck with hardware failure or power failure. So no packets can be transmitted through the dead end. We cannot pass through the dead end and the copy on that route gets struck.

In our proposed algorithm, if there is dead end we consider it in two ways. First way is to route the copy using the bypass recovery [7]. This is possible in case of route discovery. In the second way, if there is no route our algorithm's Focus phase will transmit the copy.

## 3 SIMULATION RESULTS

The simulation is done with a good simulator having four types of dynamic node formations such as 25, 50, 75,100 nodes in the network. First the dynamic nodes hop distance, is found out by the node coverage.

Then the source and the destination are identified. By this we get all the moving nodes between the source node and the destination node.

Multiple copies are sprayed into the network by spray select focus routing algorithms. Assuming a node as the source node and another node as the destination to reach, there are many moving nodes in between and we are choosing a node as moving node.

Spray Select Focus algorithm is simulated to avoid congestion and overcoming dead ends. The normal routing routes the packet to every other node near by Fig 2, so the packets are spread to all the nodes which are in neighbour to the source.

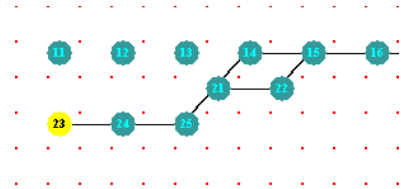

Fig. 1. Normal Routing

By our Spray Select Focus Algorithm a dead end is overcome by a Bypass recovery in the with coverage case which we can see in the following figure, Fig. 3.

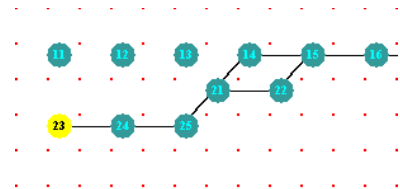

Fig. 2. Bypass Recovery

In case of without coverage the Deadend is overcome by the Focus phase of our algorithm. The following figure explains the Spray Select Focus routing Fig 4 &5 with and without coverage and dead ends.

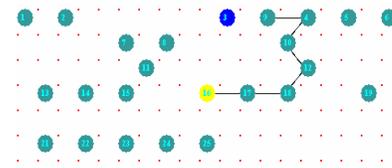

Fig.3. Spray Select Focus Without Deadends.

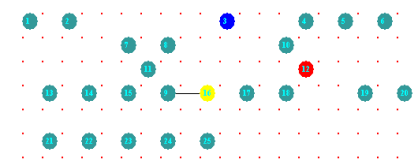

Fig.4. Spray Select Focus With Deadends.

We are going to compare the Spray and Focus algorithm which has no contention and dead ends with our Spray Select Focus algorithm with contention and dead ends.

The analysis is done with the following parameters:
- Transmission Rate(TR)
- Packet Delay(PD)
- Hop Distance (H)

Transmission Rate = Number of nodes Covered (H).

PD= PS/H * T. Where PD,PS is the Packet Delay and Packet Size, H is the number of nodes covered by hops and T is the minimum time taken to deliver the packets.

H= Distance between source and destination

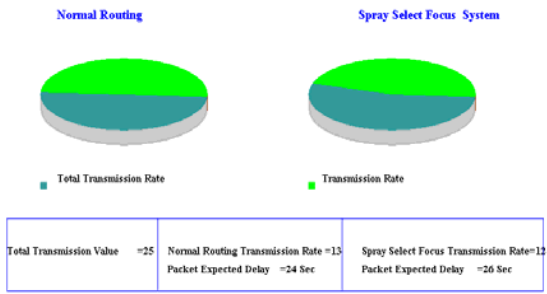

Fig. 5. Transmission rate for Normal and Spray Select Focus Routing without deadends

From the above figure we can say we have close to optimal transmission rate than Normal routing is shown in the above graph Fig 6. We have done analysis with our Spray Select Focus Routing with and without dead ends in the network. Fig 7 shows the transmission rate for Normal and Spray Select Focus Routing with Deadends.

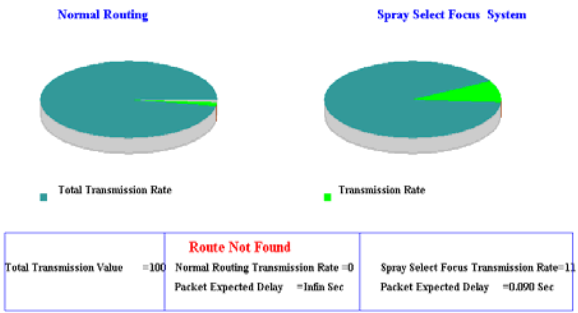

Fig. 7. Transmission rate for Normal and Spray Select Focus Routing with deadends.

Then we have done an analysis with Packet delay with the packet sizes 5,10,15,20 & 25. Fig 8 & 9 shows the bar chart for Packet delays with and without deadends.

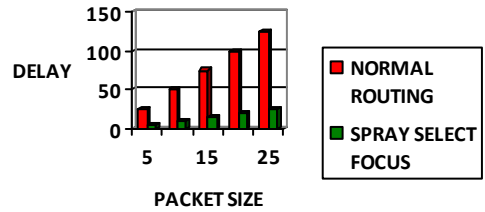

Fig. 8. Packet delay for both routings without deadends.

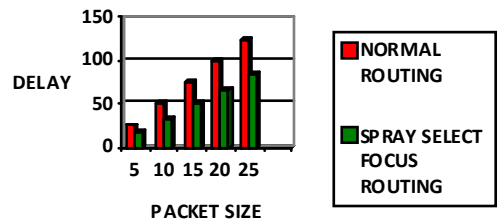

Fig. 9. Packet delay for both routings with deadends

The above figure shows the high Packet delay for normal routing in the cases of with and without dead ends in the network. The Hop distance is calculated for both the cases and the following graphs are drawn, Fig 10 & 11.

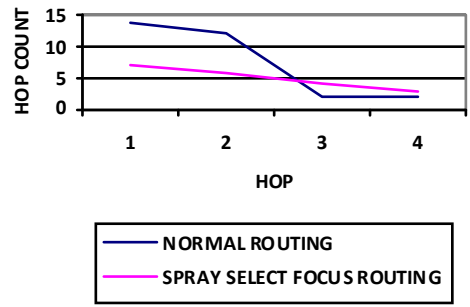

Fig. 10. Hop Distance For Normal and Spray Select Focus Routing With Deadends

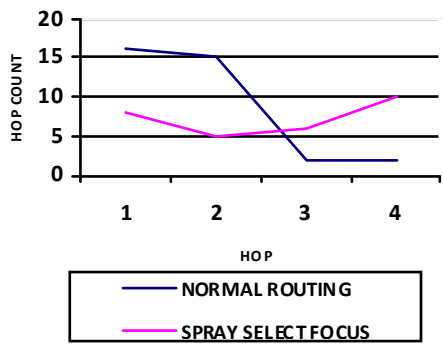

Fig. 11. Hop Distance For Normal and Spray Select Focus Routing Without Deadends.

Here we have a small and optimized hop distance in our proposed algorithm and in Normal routing it has a larger hop distance.

## 4 CONCLUSION

Routing multiple copies in a mobile ad hoc network forwarded from the source to the destination does not have end to end paths. In this work, the investigation is about the problem of multi-copy routing in mobile ad hoc networks. Spray Select Focus algorithm is used for avoiding contention and bypass recovery for dead ends.

Through simulations we have shown that avoiding dead ends and congestion leads to larger delays. Our Algorithm works well in With and Without Coverage's and our Focus phase worked well in the case of without coverage – with deadends and reduced contention. So to conclude we say that considering contention and dead ends is very important and it leads to inaccurate results. And we say that our algorithm performed well than the previous routing schemes in case of with and without coverage and with and without dead ends. Finally it is robust in case of delivering the message to the destination.

**Ms.E. Jenefa JebaJothi** has received his B.Sc degree in Computer science in 2004 from MS University and MCA degree in 2007 from MS University, Tirunelveli, Tamilnadu, India. She is now doing M.E in Anna University Tirunelveli, Tamilnadu, India. Her areas of interest are Ad hoc Networks, Network security and software Engineer-ing. She has presented many papers in national and International Conferences. As part of this paper, she is working on developing routing protocols for mobile ad hoc networks. She is a member of IAENG.

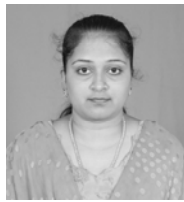

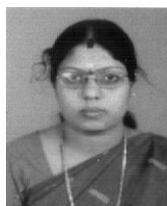

**Dr.V.Kavitha** obtained her B.E degree in Computer Science and Engg in 1996 from MS University and ME degree in Computer Science and Engineering in 2000 from Madurai KamaRaj University. She is the University Rank Holder in UG and Gold Medalist in PG. She received PhD degree in Computer Science and Engineering from Anna University Chennai in 2009. Right from 1996 she is in the Department of Computer Science & Engineering under various designations. Presently she is working as Asst. Prof in the Department of CSE at Anna University Tirunelveli. In addition She is the Director In-Charge of University V.O.C College of Engineering. Tuticorin. Currently, under her guidance ten Research Scholars are pursuing PhD as full time and part time. Her research interests are Wireless networks Mobile Computing, Network Security, Wireless Sensor Networks, Image Processing and Cloud Computing. She has published many papers in national and International journal in areas such as Network security, Mobile Computing, wireless network security, and Cloud Computing. She is a life time member of ISTE.

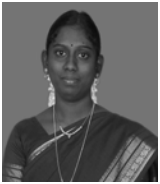

**Ms. T. Kavitha** obtained her B.E degree in Computer Science and Engineering from MK University in 1999 and M.E degree in Govt. college of Engineering, Tirunelveli, Tamilnadu. Right from 2005 she is in the Department of Computer Science & Engineering under various designations. She is now working as a Lecturer in the Department of CSE at Anna University Tirunelveli, Tirunelveli, Tamilnadu, India. Her research interests are Wireless Sensor Networks, Mobile Ad hoc Networks, and Image Processing. She has presented many papers in National and International Conferences in the same area. She is a member of ISTE.